\documentclass[12pt]{revtex4}

\usepackage{amssymb}
\usepackage{amsfonts}
\usepackage{amsmath}
\usepackage[latin1]{inputenc}
\usepackage{graphicx,color}

\makeatletter

\newcommand{\ep}{\epsilon}

\newcommand{\be}{\begin{equation}}
\newcommand{\ee}{\end{equation}}
\newcommand{\bea}{\begin{eqnarray}}
\newcommand{\eea}{\end{eqnarray}}

\begin{document}

\title{Duality invariance implies Poincar\'e invariance}
\author{Claudio Bunster$^{1,2}$ and Marc Henneaux$^{1,3}$}
\affiliation{${}^1$Centro de Estudios Cient\'{\i}ficos (CECs), Casilla 1469, Valdivia, Chile}
\affiliation{${}^2$Universidad Andr\'es Bello, Av. Rep\'ublica 440, Santiago, Chile} 
\affiliation{${}^3$Universit\'e Libre de Bruxelles and International Solvay Institutes, ULB-Campus Plaine CP231, B-1050 Brussels, Belgium}

\begin{abstract}
We consider  all  possible dynamical theories which evolve two transverse vector fields out of a three-dimensional Euclidean hyperplane, subject to only two assumptions: (i) the evolution is  local in space, and (ii) the theory is invariant under ``duality rotations" of the vector fields into one another. The commutators of the Hamiltonian and momentum densities are shown to be necessarily those of the Poincar\' e group or its zero signature contraction. Space-time structure thus emerges out of  the principle of duality.
\end{abstract}

\maketitle

The original description of electricity and magnetism devised by Faraday was formulated in terms of electric and magnetic lines of force.  In its simplest and purest form, the lines of force never end.   There are no sources. In contemporary mathematical parlance, the electric and magnetic fields are divergence-free. In its most generic formulation, the principle of electric-magnetic duality states that the electric field density ${\mathcal E}^i$ and magnetic field density ${\mathcal B}^i$ are to be treated on the same footing. It is then natural to introduce a notation which enables one to easily keep track of their interchange.  One thus writes,
\be
({\mathcal B}^{\; i}_a )= ({\mathcal B}^i , {\mathcal E}^i ), \; \; \; a=1,2,
\ee
and gives the following precise meaning to the expression ``on the same footing":  one demands rotational invariance in the two-dimensional plane whose axes are labelled by the index $a$.  These rotations are termed ``duality rotations".

If one demands that the ${\mathcal B}^{\; i}_a$ should have a dynamical evolution with a Hamiltonian structure, one needs to introduce a Poisson bracket among them.    The simplest possibility that is invariant under duality rotations and spatial rotations, is local in space, and is consistent with the divergence-free character,
\be
{\mathcal B}^{\; i}_{a\; , i} = 0
\ee
of ${\mathcal B}^{\; i}_a$ is,
\be
[{\mathcal B}^{\; i}_a(x) , {\mathcal B}^{\; j}_b(x')] = - \ep^{ijk} \ep_{ab} \delta_{,k}(x,x'), \label{BracketEB}
\ee
and it is actually unique \cite{unique}.
It follows from (\ref{BracketEB}) that the Poisson bracket of two duality-invariant quantities is duality invariant.

We will call ``local" an expression where the  ${\mathcal B}^{\; i}_a$ appear undifferentiated (sometimes this very restricted notion is called ``ultralocal" because no derivatives are admitted). Two duality invariant quantities that will play a key role in what follows are the scalar
\be
h = \frac12 {\mathcal B}^{\; i}_a {\mathcal B}^{\; j}_b \delta^{ab} \delta_{ij} \label{Defh}
\ee
and the vector 
\be
{\mathcal H}_k = - \frac12 {\mathcal B}^{\; i}_a {\mathcal B}^{\; j}_b \ep^{ab} \ep_{ijk}.
\ee

From the bracket (\ref{BracketEB}), one may immediately verify that the Lie derivative of any functional $F[{\mathcal B}^{\; i}_a]$ along a spatial vector field $\xi^i(x)$ is given by 
\be
{\mathcal L}_\xi F = [F,  \int d^3x \, \xi^i(x) {\mathcal H}_i(x)].
\ee
This means that  
$
{\mathcal H}_i(x) $ is the momentum density in curvilinear coordinates. Thus, in particular,  in Cartesian coordinates ${\mathcal H}_i(x)$ is the linear momentum density while in cylindrical coordinates ${\mathcal H}_\varphi(x)$ is the angular momentum density around the $z$-axis.  The ${\mathcal H}_i(x)$ obey the Poisson bracket algebra
\be
[{\mathcal H}_i(x), {\mathcal H}_j(x')] = {\mathcal H}_i(x') \delta_{,j}(x,x') + {\mathcal H}_j(x) \delta_{,i}(x,x'). \label{BracketHiHj}
\ee

In order to introduce dynamics, we need to bring in a Hamitonian $H$, which will evolve the fields off a given initial three-dimensional surface. We will demand that it be of the form 
\be  
H = \int d^3x  \, {\mathcal H}(x),
\ee
where ${\mathcal H}(x)$ is a local duality and rotation invariant function constructed out of the ${\mathcal B}^{\; i}_a$.  We will also require that the complete set ${\mathcal H}(x)$, ${\mathcal H}_i(x)$ forms an algebra of which (\ref{BracketHiHj})  is a subalgebra. 

With these requirements only, we will prove further below that,
\be
[{\mathcal H}(x), {\mathcal H}(x')] = - \epsilon \, \delta^{ij}\left({\mathcal H}_i(x')  + {\mathcal H}_i(x)\right) \delta_{,j}(x,x') \label{BracketHH}
\ee
where $\epsilon = 0$ or $-1$.
Once this equation is established, we have proven our point.  Space-time invariance emerges out of duality invariance. 

Indeed, for $\epsilon = -1$, Eq. (\ref{BracketHH}) is precisely the commutation rule for the energy densities shown in \cite{Dirac:1962aa,Schwinger:1963xx} to be the condition for a field theory to be Poincar\'e invariant (see also \cite{Teitelboim:1972vw}).  This equation was referred to in the concluding sentence of \cite{Schwinger:1963xx} as {\it ``what may well be considered the most fundamental equation of relativistic quantum field theory"}.

The case $\epsilon = 0$ has been termed the ``zero-signature" case \cite{CTZeroSignature}.  It corresponds  to a spacetime geometry  whose invariance group is the contraction of the Poincar\'e group when the speed of light goes to zero \cite{MH0}.  This Carroll group ({\it ``Now, here, you see, it takes all the running you can do, to keep in the same place"}) was first encountered in a systematic study of possible extensions of the three-dimensional Euclidean group \cite{BacryLL} .  The zero signature geometry finds an interesting application in connection with the decoupling of spatial points
near the generic singularity in the early universe \cite{BKL,Damour:2002et}.  It also has been used as the starting point, corresponding to the ``free case", for a perturbation theory in quantum gravity \cite{Teitelboim:1981ua}.

It is quite remarkable that SO(2) duality rotations, which are a {\em circular} Euclidean invariance, give raise in spacetime to {\em hyperbolic} Lorentz invariance.  Conversely, if one changes the circular $\delta _{ab}$ by the hyperbolic $\eta_{ ab}= diag(-1,1)$, all the analysis in this letter could be repeated and one would arrive at $\epsilon= +1$ in (\ref{BracketHH}), corresponding to Euclidean spacetime. Thus one sees another fascinating imprint of duality in spacetime structure. Performing a Wick rotation $\alpha \rightarrow i\alpha$ of the duality angle implies a Wick rotation of time
$x^0 \rightarrow i x^0$.

The technical steps of the proof of (\ref{BracketHH}) are to a considerable extent given in \cite{Deser:1997gq} where a different problem was treated. There, {\em both (\ref{BracketHH}) and duality invariance} were imposed to obtain restrictions on the form of ${\mathcal H}$. The key difference with the present work is that the commutation relations (\ref{BracketHH}) {\em do not have to be assumed independently}, but rather are {\em implied by duality invariance.} A fortiori, the restrictions on the possible ${\mathcal H}$'s follow therefore {\em from duality invariance} alone.  Every duality-invariant theory is relativistic.

The proof of (\ref{BracketHH}) goes as follows. First, one observes that since ${\mathcal H}(x)$ is a local duality-invariant function, it depends only on the two invariants $h$ given by (\ref{Defh}) and $v$ defined by
\be v =  {\mathcal H}_k {\mathcal H}^k,
\ee
Hence, $
{\mathcal H}= f(h,v).
$
Now, the brackets between $h$ and $v$ that follow from the basic brackets (\ref{BracketEB}) read
\bea
&& [h(x), h(x') ] =  \delta^{ij}\left({\mathcal H}_i(x')  + {\mathcal H}_i(x)\right) \delta_{,j}(x,x')\\
&& [h(x), v(x') ] =   2 \delta^{ij}({h(x') \mathcal H}_i(x')  \nonumber \\
&& \hspace{1.5cm} + h(x) {\mathcal H}_i(x)) \delta_{,j}(x,x') +  2 {\mathcal H}^k_{\; ,k} h \delta(x,x')\\
&& [v(x), v(x') ]=  4 \delta^{ij}\left(v(x') {\mathcal H}_i(x')  + v(x) {\mathcal H}_i(x)\right) \delta_{,j}(x,x')\nonumber \\
\eea
This implies that the bracket $[{\mathcal H}(x), {\mathcal H}(x')]$ itself is given by
\be
[{\mathcal H}(x), {\mathcal H}(x')] =  \delta^{ij}\left(F(x') {\mathcal H}_i(x')  + F(x) {\mathcal H}_i(x)\right) \delta_{,j}(x,x') \label{14}
\ee
where $F$ is equal to 
\be
F = \left(f_h\right)^2 + 4 h f_h f_v + 4 v \left(f_v\right)^2.
\ee
Here, $f_h$ and $f_v$ denotes the partial derivatives of $f$ with respect to $h$ and $v$, respectively.  

We thus see that the mere requirements of duality invariance, rotation invariance and locality imply that the bracket $[{\mathcal H}(x), {\mathcal H}(x')] $ necessarily has the form (\ref{BracketHH}), but with an overall coefficient $F$ which can be at this stage a function of the dynamical variables. 

If we now implement the additional condition that ${\mathcal H}(x)$ and ${\mathcal H}_k(x)$ should form an algebra \cite{Algebra}, we must require that $F$ be a constant.  This yields the differential equation 
\be 
\left(f_h\right)^2 + 4 h f_h f_v + 4 v \left(f_v\right)^2 = k \label{key}
\ee
for the unknown function $f$, where $k$ is a constant. 

To complete the proof of our claim, we observe that the constant $k$ is non negative.  Indeed, its value (which does not depend on $h$ or $v$ since it is a constant) can be evaluated at zero values of the fields.  For $u=v=0$, the equation (\ref{key}) reduces to $k = \left(f_h\right)^2$, which manifestly shows that $k$ is non negative.  If $k= 0$, the resulting algebra is the zero-signature algebra $\ep=0$.  If $k>0$, one can set $k=1$ by rescaling the generators and one gets the algebra (\ref{BracketHH}) with $\ep = -1$ \cite{HigherD}.

Lastly, a few comments about the solutions of (\ref{key}).  For $\epsilon= -1$ ($ k$ positive) they have been studied extensively in \cite{Gibbons:1995cv,Deser:1997gq}. By introducing the variable $s$ through the relation
$$ h=h, \; \; \; s^2 = h^2 - v \geq\left( ({\mathcal E})^2 - ({\mathcal B})^2\right)^2 \geq 0, $$ and defining further $$h = U + V, \; \; \; \; s = U-V, $$ the equation (\ref{key}) can be cast in the simple form
\be 
f_U \, f_V = k, \label{17}
\ee
which is the Hamilton-Jacobi equation in light like coordinates for a massive particle in two dimensions
For $\epsilon= -1$,  two well-known interesting solutions in closed form are $f = h $
corresponding to the standard Maxwell theory,  and
$f = \sqrt{1 + 2 h + v} $ corresponding to the Born-Infeld theory.  Equation (\ref{17}) was studied in its own right, independently of any consideration about electric-magnetic duality in \cite{Courant}.

In the Carroll case, the mass vanishes and the equation reduces to
\be 
f_U \, f_V = 0, \label{keyBis}
\ee
whose general solutions are either functions of $U$ or functions of $V$.  

We believe that the argument presented in this letter reinforces the view that there is a profound connection between duality invariance and spacetime structure, and that the latter may even emerge from the former.  

\section*{Acknowledgments} 
Useful comments by C. Bachas, B. Julia and H. Nicolai are gratefully acknowledged.
 Both authors  thank  the Alexander von Humboldt Foundation for Humboldt Research Awards.  They are also grateful to the Max Planck Institute for Gravitational Physics (Albert Einstein Institute) for kind hospitality while this work was being carried out. The work of M.H. is partially supported by the ERC through the ``SyDuGraM" Advanced Grant, by IISN - Belgium (conventions 4.4511.06 and 4.4514.08) and by the ``Communaut\'e Fran\c{c}aise de Belgique" through the ARC program.  The Centro de Estudios Cient\'{\i}ficos (CECS) is funded by the Chilean Government through the Centers of Excellence Base Financing Program of Conicyt.


\begin{thebibliography}{99}
\bibitem{unique} An illuminating way of seeing the uniqueness of the bracket (\ref{BracketEB}) is the following. The divergence-free character of ${\mathcal B}^i$ implies the existence of a vector-potential $A_i$ which is determined up to the addition of a gradient (gauge transformation).  One then introduces the Hamiltonian structure through a gauge-invariant Lagrangian.  We call electric field  ${\mathcal E}^i$ the negative of the canonically conjugate momentum  to $A_i$.    It is then always divergence-free and Eq. (\ref{BracketEB}) holds.  Note that although for Maxwell theory ${\mathcal E}^i = F^{0i}$, this will not be the case for more general Lagrangians such as the one of Born and Infeld. In general, $F^{0i}$ is not divergence-free  and hence it does not admit a description in terms of lines of force, although ${\mathcal E}^i$ always does. A more common and proper terminology for ${\mathcal E}^i$ would in fact be ``electric displacement field", but we use the words ``electric field" for short since there will be no risk of confusion.
\end{thebibliography}
\end{document}